\documentclass[a4paper,11pt]{article}
\usepackage{pos}
\usepackage{array} 
\usepackage{float}
\usepackage{hyperref}
\usepackage[dvipsnames]{xcolor}
\usepackage{textcomp}

\definecolor{pomegranate_purple}{HTML}{b90078}

\hypersetup{
    colorlinks   = true,
    citecolor    = pomegranate_purple,
    urlcolor     = pomegranate_purple,
    linkcolor    = pomegranate_purple
}

\title{The Dark Matters at ICRC 2025}
\ShortTitle{DM Rapporteur ICRC 2025}
\author*[a]{Milena Crnogor\v{c}evi\'{c}}
\affiliation[a]{Stockholm University and The Oskar Klein Centre for Cosmoparticle Physics,\\
  Alba Nova, 10691 Stockholm, Sweden}
\emailAdd{milena.crnogorcevic@fysik.su.se}
\abstract{The dark matter track at ICRC~2025 showed a field in transition. Direct detection has entered the \emph{neutrino-floor era}, with XENONnT and PandaX-4T now limited by Solar neutrinos. Indirect searches have become truly \emph{multimessenger}, combining $\gamma$-rays, neutrinos, cosmic rays, and radio data under unified likelihoods and shared systematics. Non-WIMP candidates---axions, sub-GeV particles, primordial black holes, macroscopic relics---are becoming central. Across all fronts, progress depends as much on new detectors as on the coherence of shared data, methods, and analysis frameworks. Here, I distill the main experimental and conceptual shifts behind these trends, noting how assumptions have evolved since ICRC~2023 and where the next decisive advances are likely to come.}
\ConferenceLogo{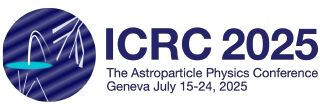}
\FullConference{39th International Cosmic Ray Conference (ICRC2025)\\
 15–24 July 2025\\
Geneva, Switzerland\\}
\begin{document}
\maketitle

\section{Introduction}

``What is essential is invisible to the eye,'' said the Little Prince. At ICRC 2025, close to eighty contributions proved that dark matter still fits this description perfectly. After four decades of increasingly sophisticated searches---spanning underground laboratories, space- and ground-based telescopes, and neutrino observatories---the field has compiled a record of steadily tightening constraints, but no detection\footnote{For comprehensive reviews of dark matter see, e.g.,  \cite{Bertone:2010, Bertone:2016, Bauer:2017, Profumo:2017, Marsh:2024}. Many more exist. From here on, I reference only ICRC 2025 proceedings due to space constraints.}.

The ICRC 2025 showcased this evolution through 78 dark matter contributions (61 talks, 14 posters, and 3 plenaries), spanning ninety orders of magnitude in mass: from ultralight axion-like particles (ALPs), through sub-GeV and MeV-scale candidates, the electroweak weakly interacting massive particles (WIMPs), heavy and superheavy relics, up to primordial black holes (PBHs) and macroscopic objects. On the experimental front, searches employed a similarly diverse set of messengers: $\gamma$-rays, neutrinos, cosmic-ray antimatter, radio emission, and direct detection. The breadth in theoretical models and detection techniques was matched by the analytical maturity: multi-instrument joint likelihoods are becoming standard in dwarf-galaxy searches, multimessenger strategies increasingly define the state of the art in indirect detection, and machine learning techniques now enhance  background discrimination in direct detection experiments.

\begin{figure}[H]
\centering
\includegraphics[width=\textwidth]{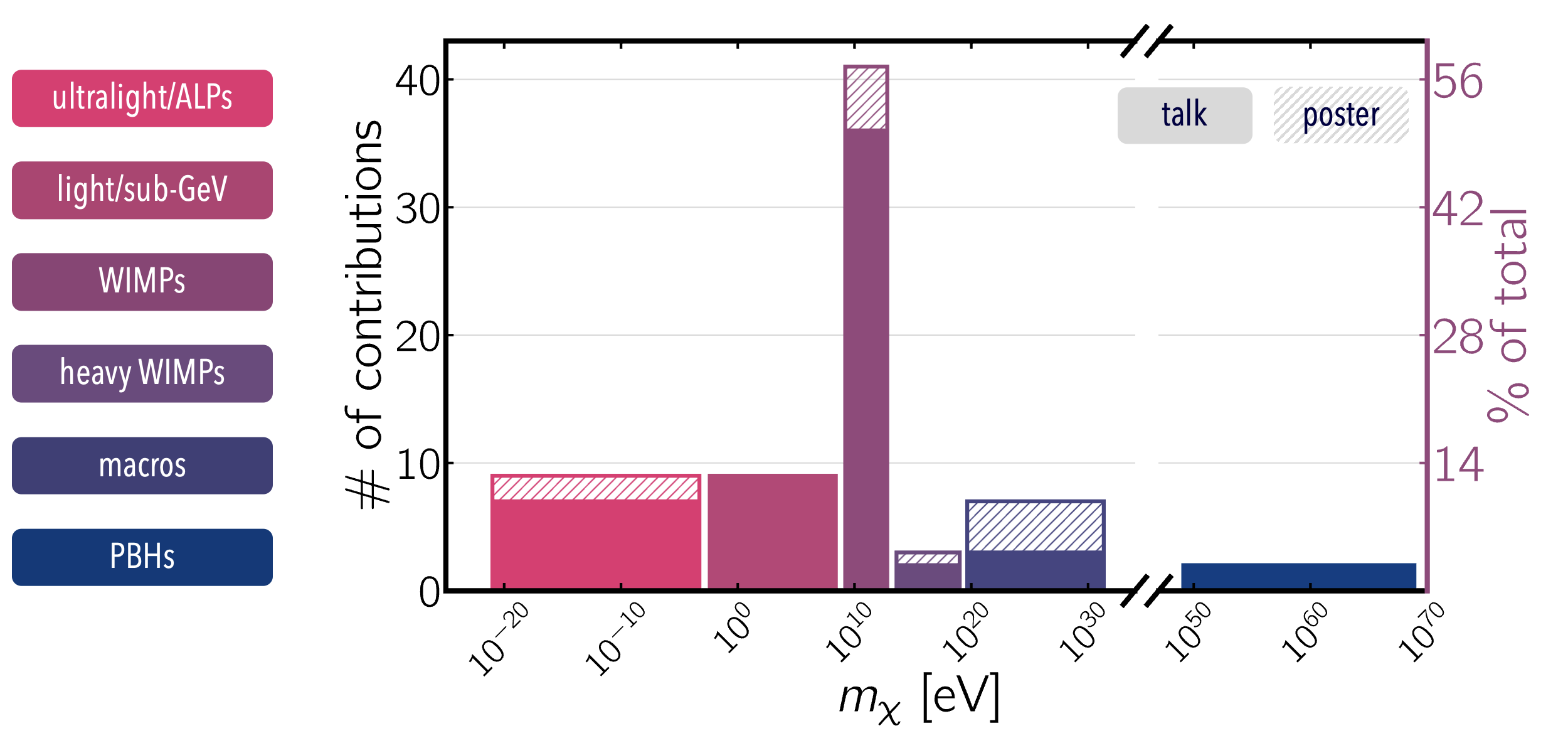}
\caption{Distribution of dark matter ICRC 2025 contributions by dark matter particle mass. The bar height represents the number of contributions (left $y$-axis), with the corresponding percentage of total contributions shown on the right $y$-axis. Filled segments indicate talks and dashed segments represent posters.}
\label{fig:breakdown}
\end{figure}

Compared with ICRC 2023, four shifts stand out. \textbf{First}, direct detection has entered the \emph{neutrino floor era}. XENONnT and PandaX-4T now detect Solar neutrinos directly, confirming that the dominant background for WIMP searches is no longer instrumental noise but irreducible astrophysical signals.
\textbf{Second}, indirect searches increasingly adopt \emph{joint multi-instrument likelihoods} with consistent $J$-factor priors and shared systematics, integrating complementary datasets under unified statistical frameworks. The first comprehensive dwarf-galaxy analysis combined \textit{Fermi}-LAT, HAWC, H.E.S.S., MAGIC, and VERITAS data---spanning 100 MeV to beyond 10 TeV in energy coverage---and improved sensitivity by factors of 2--3 relative to any single instrument.
\textbf{Third}, non-WIMP candidates have moved from periphery to center: for the first time in recent ICRC history, the plenary program devoted more space to non-WIMP scenarios than to canonical WIMPs, reflecting the field's strategic broadening towards alternative dark matter candidates.
\textbf{Fourth}, and perhaps most notably, ICRC 2025 revealed a structural convergence across traditionally separate communities. Astrophysical neutrino detectors---such as IceCube---are repurposed as precision dark matter observatories, while direct detection experiments now reach the sensitivity required to measure the Solar neutrinos. This convergence underscores an emerging reality: while next-generation detectors remain essential for exploring new parameter space, near-term progress increasingly depends on collaborative infrastructure---shared analysis tools, open data, open-source software---that enables coherent integration of diverse datasets. Given present funding constraints and uncertain timelines for next-generation experiments, maximizing scientific return from existing data has become not merely advantageous but essential.

This rapporteur paper is inevitably subjective and deliberately selective. Rather than cataloging results experiment by experiment, I aim to \textbf{underpin structural lessons} that emerge when contributions are viewed together. My aim is to capture the field's status as of 2025: to identify which assumptions have shifted since ICRC 2023, which methods now define the state of the art, and where the next decade of searches is likely to be decisive. If the last decade was defined by steady exclusion driven largely by increasing exposure and improved control of systematics, the present moment marks a transition: dark matter searches now depend as much on coherent, multi-instrument analysis as on continued advances in individual sensitivity.


To reflect these shifts, the organization of this paper therefore follows physics themes rather than instrumental categories. Section~\ref{sec:direct} reviews the evolution of direct detection and its approach to the neutrino floor. Section~\ref{sec:indirect} surveys indirect searches organized by astrophysical target---dwarf galaxies, the Galactic Center, Galactic halo, galaxy clusters, and others---rather than by messenger or experiment, emphasizing the genuine multimessenger nature of current indirect searches. I provide my concluding reflections in Sec.~\ref{sec:concl}.

\section{Direct Detection: The Neutrino-Floor Era}
\label{sec:direct}

Direct detection has reached a defining milestone. For the first time, liquid-xenon time-projection chambers (TPCs) have become sensitive to coherent elastic neutrino-nucleus scattering (CE$\nu$NS) from solar $^8$B neutrinos---marking the transition from noise-limited to neutrino-limited searches.

\subsection{Reaching the Neutrino Floor}

At ICRC~2025, \textbf{XENONnT} reported a 2.73$\sigma$ indication of CE$\nu$NS events together with an updated WIMP limit of $\sigma_{\mathrm{SI}} < 1.7\times10^{-47}$~cm$^2$ at a dark matter mass, $m_\chi$, of 30~GeV, improving by a factor of 1.8 over SR0 \cite{XENONnT:2025}. Sub-$\mu$Bq/kg electron-recoil backgrounds, achieved through radon removal, place XENONnT operation at the $pp$-solar-neutrino background level. \textbf{PandaX-4T} independently reported a 2.64$\sigma$ $^8$B CE$\nu$NS indication in 1.54~t$\cdot$yr exposure, along with competitive constraints on axion, ALP, and dark-photon scenarios \cite{PandaX4T:2025}. Together, these results establish that, in the multi-GeV–TeV range, xenon experiments are no longer limited by instrumental noise or radiogenic contamination, but by the irreducible neutrino background itself.

With CE$\nu$NS now observed---or at least strongly indicated---the field of direct detection faces two strategic directions: (i) scaling exposure and refining background modeling to statistically separate a potential WIMP signal from CE$\nu$NS, and (ii) focusing on observables that CE$\nu$NS cannot mimic, such as Migdal-induced ionization or annual modulation. This twofold direction captured the ICRC~2025 message from large noble-liquid experiments: they now operate in the neutrino-fog regime, and future progress hinges on how we navigate---and ultimately extend sensitivity below---that limit.

\begin{figure}[H]
\centering
\includegraphics[width=\textwidth]{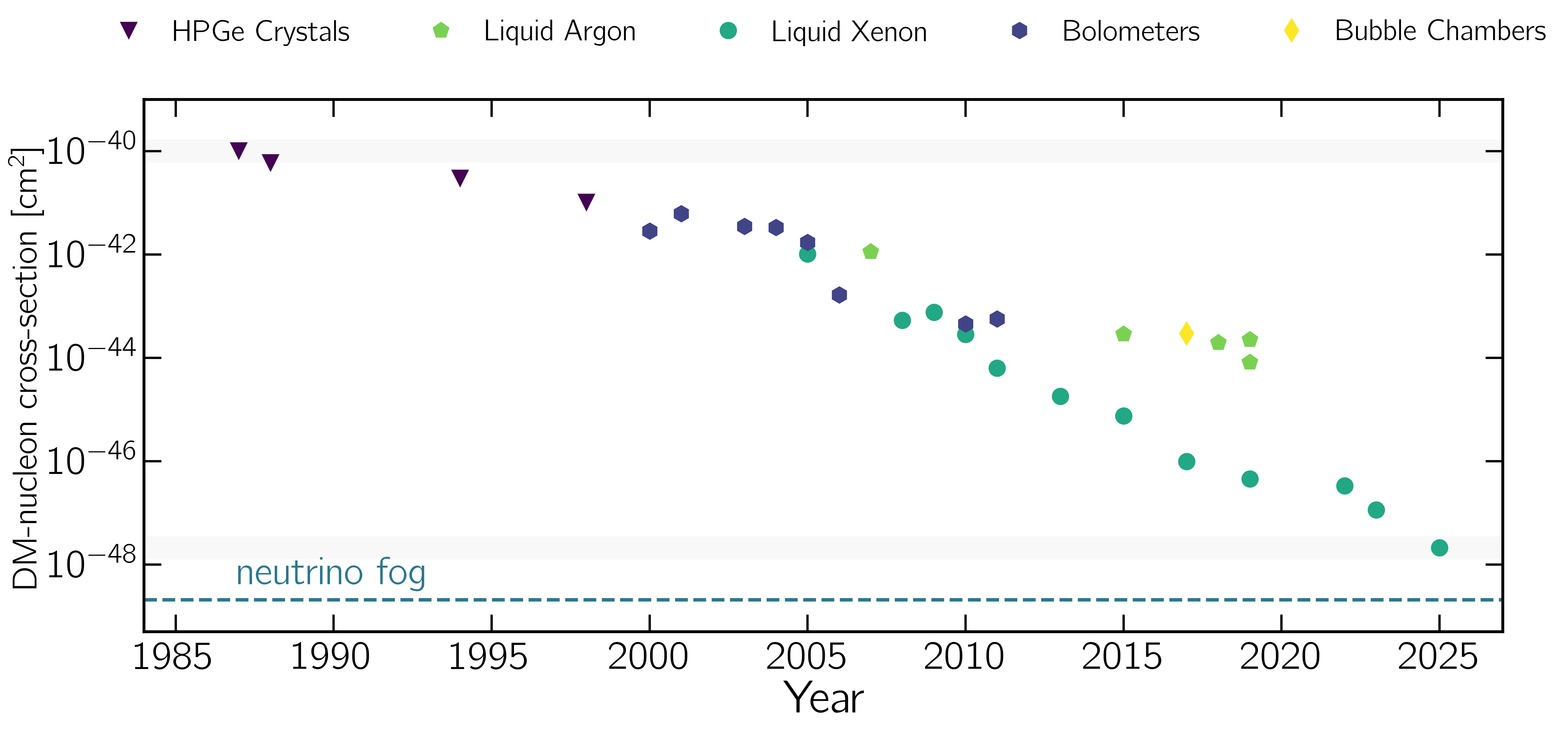}
\caption{\textbf{Four decades of progress in spin-independent WIMP searches.} Markers show the spin–independent WIMP–nucleon limits
from HPGe crystals, cryogenic bolometers, bubble chambers, liquid argon, and liquid xenon experiments for $m_\chi = 50$~GeV. The dashed line marks the onset of the \emph{neutrino fog} ($\sim$1 expected CE$\nu$NS event for multi-ton-year exposures). The two faint gray bands highlight the overall improvement of about \textbf{eight orders of magnitude in sensitivity}---from early HPGe/bolometer results to current liquid–xenon limits. Adapted from \cite{Baudis:2025}.}
\label{fig:dd_evol}
\end{figure}

Parallel progress continues in the liquid-argon program. \textbf{DarkSide-20k} is now under construction at the INFN Gran Sasso National Laboratory as the next-generation dual-phase liquid-argon TPC, with a 20~t fiducial (50~t total) target, underground argon for ultra-low backgrounds, and custom cryogenic SiPM arrays for light readout \cite{DarkSide20k:2025}. Its triggerless data acquisition system and high-rate architecture are designed to reach sensitivities at---or beyond---the neutrino floor by the end of the decade.
\textbf{DEAP-3600}, meanwhile, presented an extended 813-day dataset with an improved background model, upgraded hardware for further suppression, and neural-network-based event reconstruction. The collaboration also reported a new direct measurement of the $^{39}$Ar half-life, in mild tension with previous nuclear data \cite{Garai:2025}. 

\begin{figure}[H]
\centering
\includegraphics[width=\textwidth]{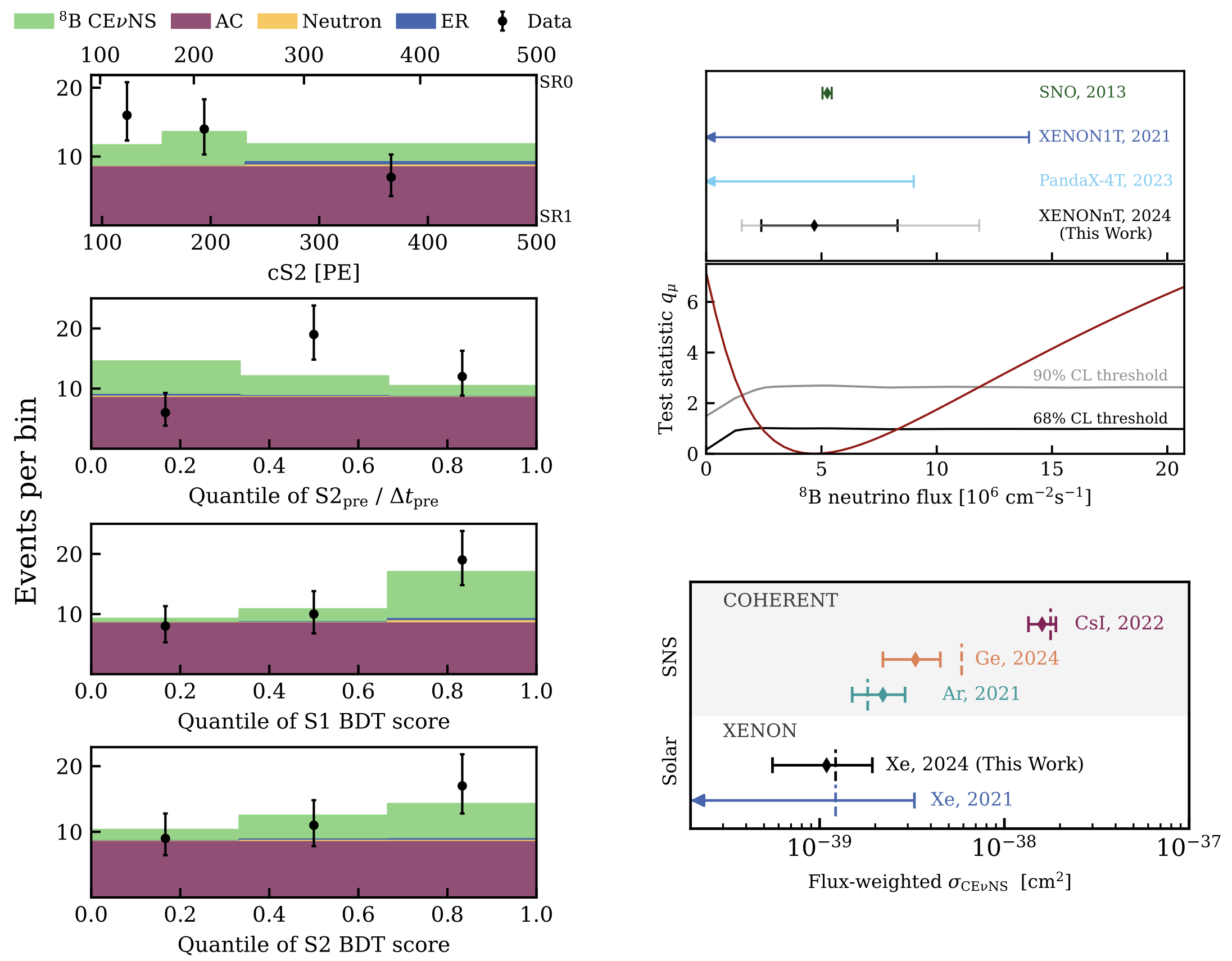}
\caption{\textbf{Observation of solar $^8$B neutrinos via CE$\nu$NS in XENONnT.} 
Left: projected distributions of best-fit signal and background components combining both science runs (SR0 and SR1). 
Data points (black) are shown with Poisson error bars; the CE$\nu$NS signal (green) is stacked above accidental coincidences (purple), electronic recoils (blue), and neutrons (yellow). 
Right: upper panel—68\% (90\%) measurement of solar $^8$B neutrino flux compared with SNO, XENON1T, and PandaX-4T (2023, see updated results in \cite{PandaX4T:2025}); lower panel—flux-weighted CE$\nu$NS cross-section measured with XENONnT, alongside COHERENT results using different targets. 
Adapted from \cite{XENONnT:2025}.}
\label{fig:xenonnt_solar}
\end{figure}

\subsection{Beyond Canonical WIMPs: Sub-GeV Dark Matter}

Over the past decade, direct detection experiments have expanded beyond the multi-GeV WIMP regime to include lighter, sub-GeV dark matter.
If dark matter never reached thermal equilibrium with the Standard Model plasma, keV--GeV masses arise naturally, producing recoil energies that are well below the sensitivity of traditional noble-liquid detectors. Probing this regime therefore requires a fundamental shift in detector technology---from measuring nuclear recoils of tens of keV to detecting ionization and phonon signals at the few-eV level. This shift has driven the development of low-threshold solid-state detectors capable of counting single charge carriers, and cryogenic crystals that measure phonons from sub-keV energy deposits. Progress in this direction was evident at ICRC~2025, where several experiments reported stable low-threshold operation and presented first results on sub-GeV dark matter searches.

\noindent\textbf{Solid-state and cryogenic detectors.} 
\textbf{SENSEI} uses silicon skipper charge-coupled devices (CCDs) with multiple non-destructive readouts per pixel, achieving sub-electron noise and single-electron sensitivity \cite{SENSEI:2025}. The experiment targets dark-matter–electron scattering, dominant for sub-GeV masses where nuclear recoils are suppressed. At ICRC~2025, the collaboration reported a single-electron rate of $(1.4\pm0.1)\times10^{-5}$~e$^-$~pix$^{-1}$~day$^{-1}$ from the second science run---the lowest ever recorded in silicon or any NIR/UV photodetector---and presented the first dedicated \emph{daily-modulation} search using data from the MINOS tunnel at Fermilab. These results extend SENSEI's sensitivity to dark matter masses of 0.5--3 MeV for direct scattering and 1--3 eV for absorption channels, constraining hidden-photon and ultralight-mediator models. A third science run with a new cryocooler and improved background control is underway. 
\textbf{SuperCDMS} employs cryogenic silicon and germanium detectors instrumented with transition-edge sensors to measure phonon and ionization signals with sub-50-eV resolution. At ICRC~2025, the collaboration reported successful commissioning of a detector tower at the Cryogenic Underground Test Facility (CUTE), demonstrating in-situ calibration and stable operation with measured baseline resolutions of 70–80~eV for germanium targets~\cite{SuperCDMS:2025}. These results validate the performance needed for the forthcoming SNOLAB installation, which will deploy 24 detectors (30~kg total) combining high-voltage phonon and phonon-charge technologies. Commissioning is scheduled for late~2025, with the first science run expected in~2026, targeting dark matter masses below~10~GeV and dark-photon absorption in the~100~eV--MeV range.

\noindent\textbf{Gaseous detectors.}
Complementary progress comes from spherical proportional counters operated with light noble gases. \textbf{NEWS-G} uses neon and helium targets to probe nuclear recoils from 0.1--1~GeV dark matter, achieving world-leading spin-dependent proton limits in this mass range~\cite{NEWSG:2025}. At ICRC~2025, the collaboration presented results from He+CH$_4$ data, reported ongoing neon analyses using MCMC techniques, and discussed developments toward the 3~m \emph{DarkSPHERE} prototype, designed to scale target mass and extend reach to lower cross sections.

\noindent\textbf{Scintillator detectors.} The \textbf{SABRE South} experiment, now under installation at the Stawell Underground Physics Laboratory (SUPL) in Australia, is part of the broader SABRE program, which also includes a Northern detector at Gran Sasso in Italy, though the two installations differ in shielding approach and deployment schedule~\cite{SABRE:2025}. SABRE South represents the first Southern Hemisphere test of the long-standing DAMA/LIBRA annual-modulation claim. It employs ultra-pure NaI(Tl) crystals with $^{\mathrm{nat}}$K concentrations of 4.3~ppb---three to eight times lower than other NaI experiments---and an active liquid-scintillator veto that suppresses the $^{40}$K background by an order of magnitude, reaching 0.72~cpd/kg/keVee. 
Located 1025~m underground at SUPL, the experiment is designed to disentangle seasonal systematics from a potential dark matter modulation. Assembly is ongoing, with commissioning planned for early~2026; discovery or exclusion of a DAMA-like signal is expected within roughly two years of operation.

These developments establish low-threshold direct detection as an integral part of the dark matter landscape, extending sensitivity to previously inaccessible sub-GeV masses.

\subsection{Shared Infrastructure for Neutrino and Dark Matter Searches}

\noindent Astrophysical neutrino detectors are being increasingly leveraged as direct dark matter observatories \cite{Locatelli:2025}. At ICRC~2025, \textbf{IceCube} presented the first dedicated sensitivity study to sub-GeV dark matter produced via cosmic-ray upscattering, demonstrating that large-volume neutrino telescopes can probe parameter space complementary to underground detectors \cite{Cavicchi:2025}. In parallel, the \textbf{RES-NOVA} collaboration reported new projections for its cryogenic PbWO$_4$ detector, originally designed for supernova neutrinos, showing competitive reach for both spin-independent and spin-dependent dark matter interactions \cite{Iachellini:2025}. The plenary talks further illustrated this cross-field trend: \textbf{FUNK} is repurposing an Auger mirror for hidden-photon searches \cite{Jaeckel:2025}, while plans are underway to establish a dark matter laboratory within the Bedretto underground facility---primarily built for earthquake research \cite{Baudis:2025}.

\begin{figure}[H]
\centering
\includegraphics[width=0.48\textwidth]{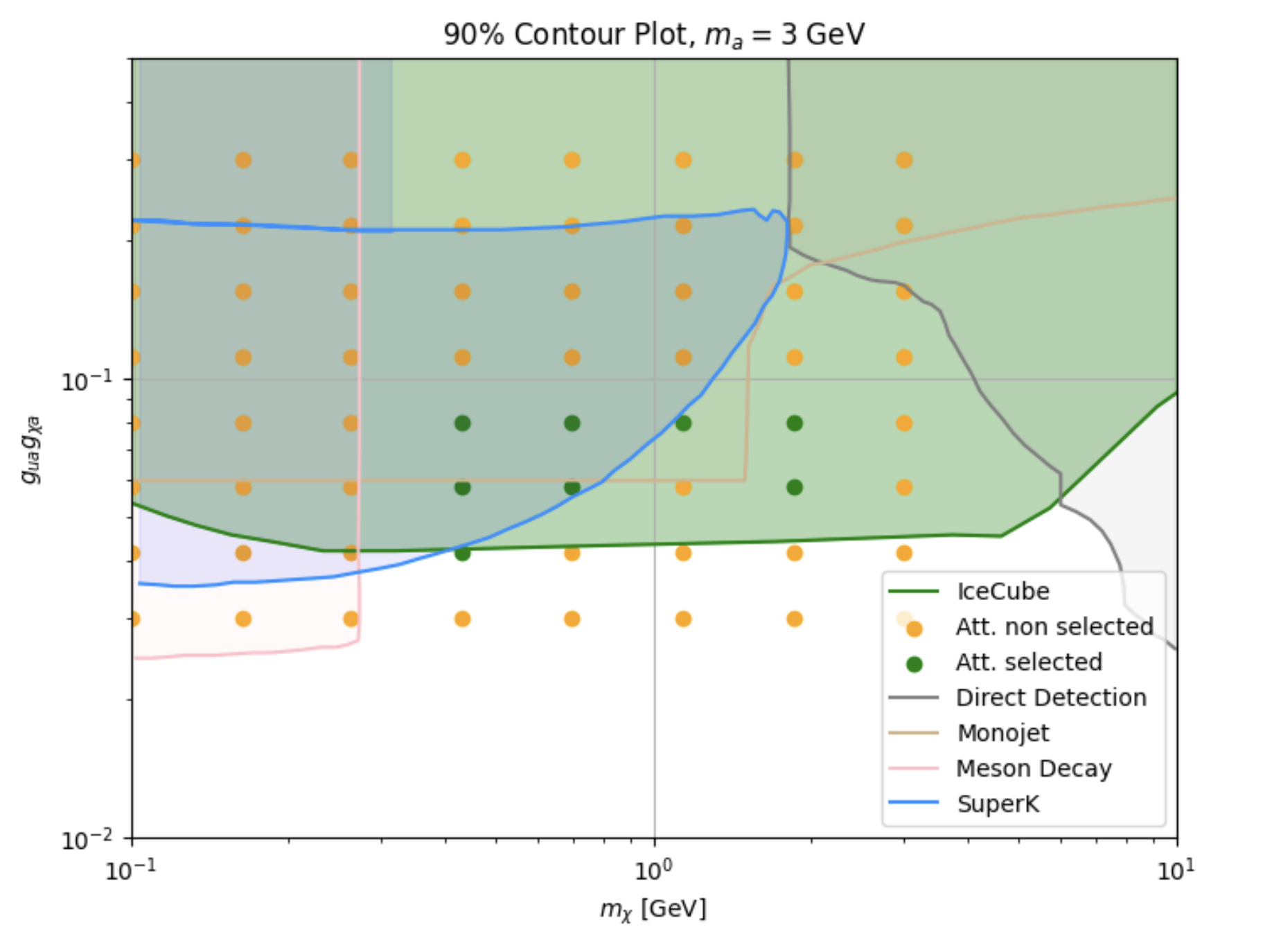}
\includegraphics[width=0.48\textwidth]{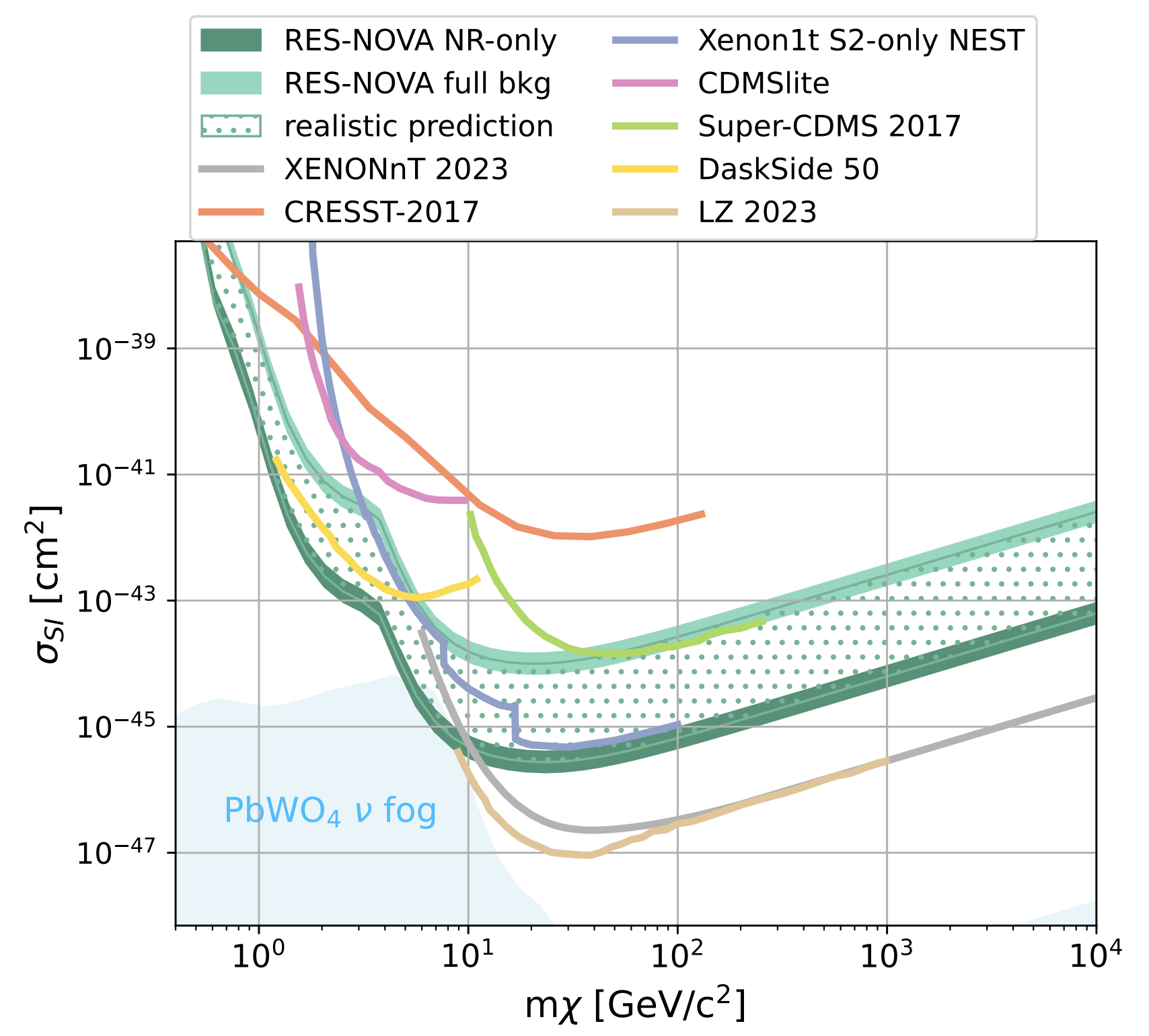}
\caption{\textbf{Complementary sensitivity of neutrino detectors to dark matter.} 
Left: IceCube 90\%~C.L.\ sensitivity to cosmic-ray upscattered sub-GeV dark matter for pseudoscalar mediators, showing complementarity with Super-Kamiokande~\cite{Cavicchi:2025}.
Right: Projected RES-NOVA sensitivity to spin-independent DM-nucleon scattering, compared with direct detection and neutrino limits~\cite{Iachellini:2025}.}
\label{fig:resnova_icecube}
\end{figure}

These developments \textbf{exemplify a broader structural convergence}: the large volume of detectors optimized for neutrino searches can be repurposed for  dark matter discovery, and vice versa. This, in my opinion, reflects more than a technical overlap---it is a pragmatic response to a tightening funding landscape, where sustained progress increasingly depends on shared infrastructure, joint analyses, and creative use of existing facilities. While \textbf{new experiments remain essential} to reach unexplored regimes, I believe the next few years \textbf{may push us to use existing instruments more effectively}---to extract new science not by expanding scale, but by expanding scope. Amid funding pressures and shifting priorities, this evolution feels not just pragmatic but necessary.

\section{Indirect Detection: The Multimessenger Era}
\label{sec:indirect}

\begin{figure}[t]
\centering
\includegraphics[width=\textwidth]{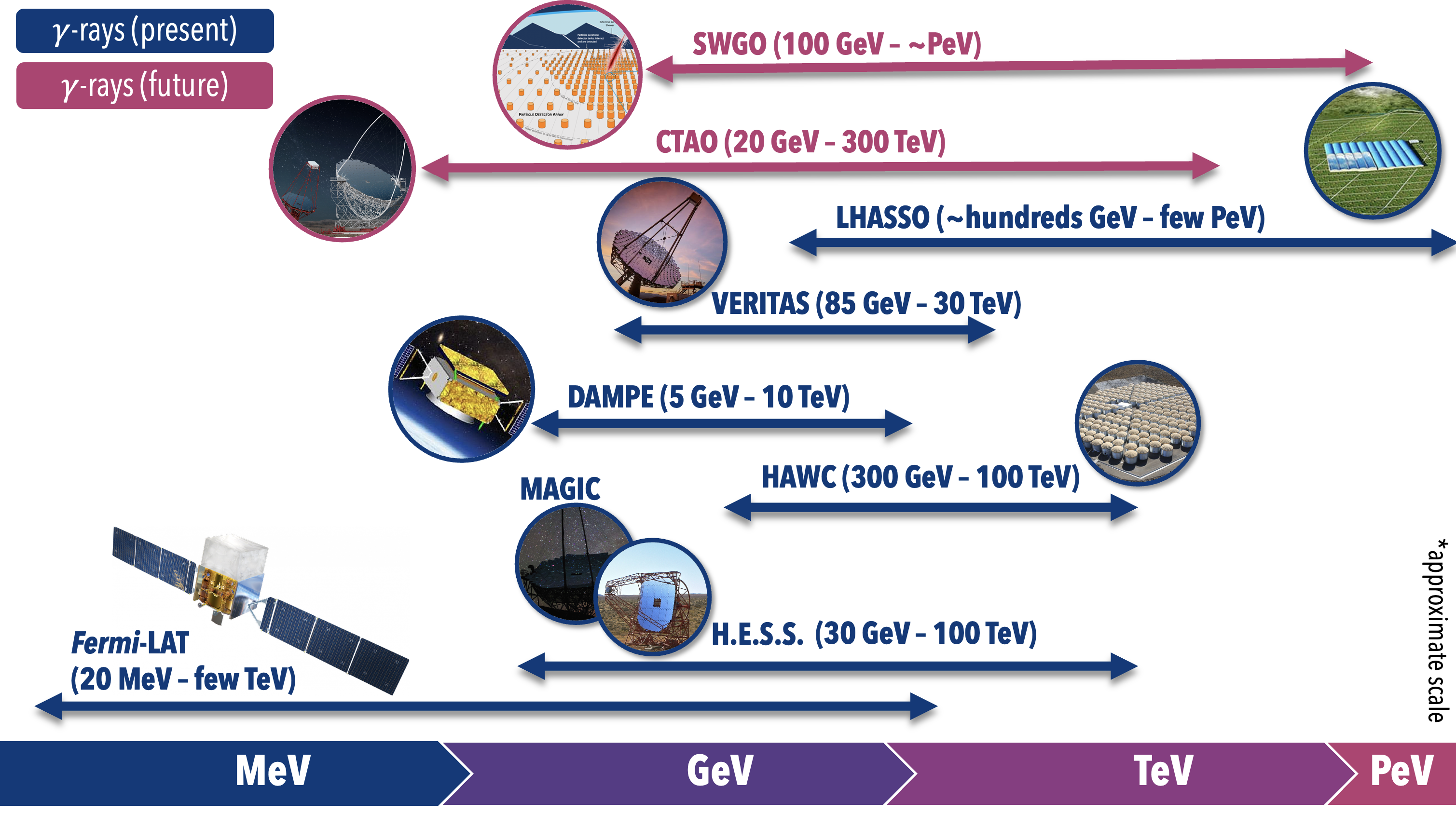}
\caption{\textbf{$\gamma$-ray coverage of current and upcoming observatories presented at ICRC~2025.} Present instruments (blue) and planned facilities (magenta) together provide continuous sensitivity from tens of MeV to multi-PeV energies. Notably, \textbf{no next-generation space-based $\gamma$-ray mission from ESA or NASA is currently planned} to succeed \textit{Fermi}, now operating beyond its expected lifetime, likely leaving a gap in MeV–GeV coverage in the coming decade.}
\label{fig:gamma-rays}
\end{figure}

\begin{figure}
\centering
\includegraphics[width=\textwidth]{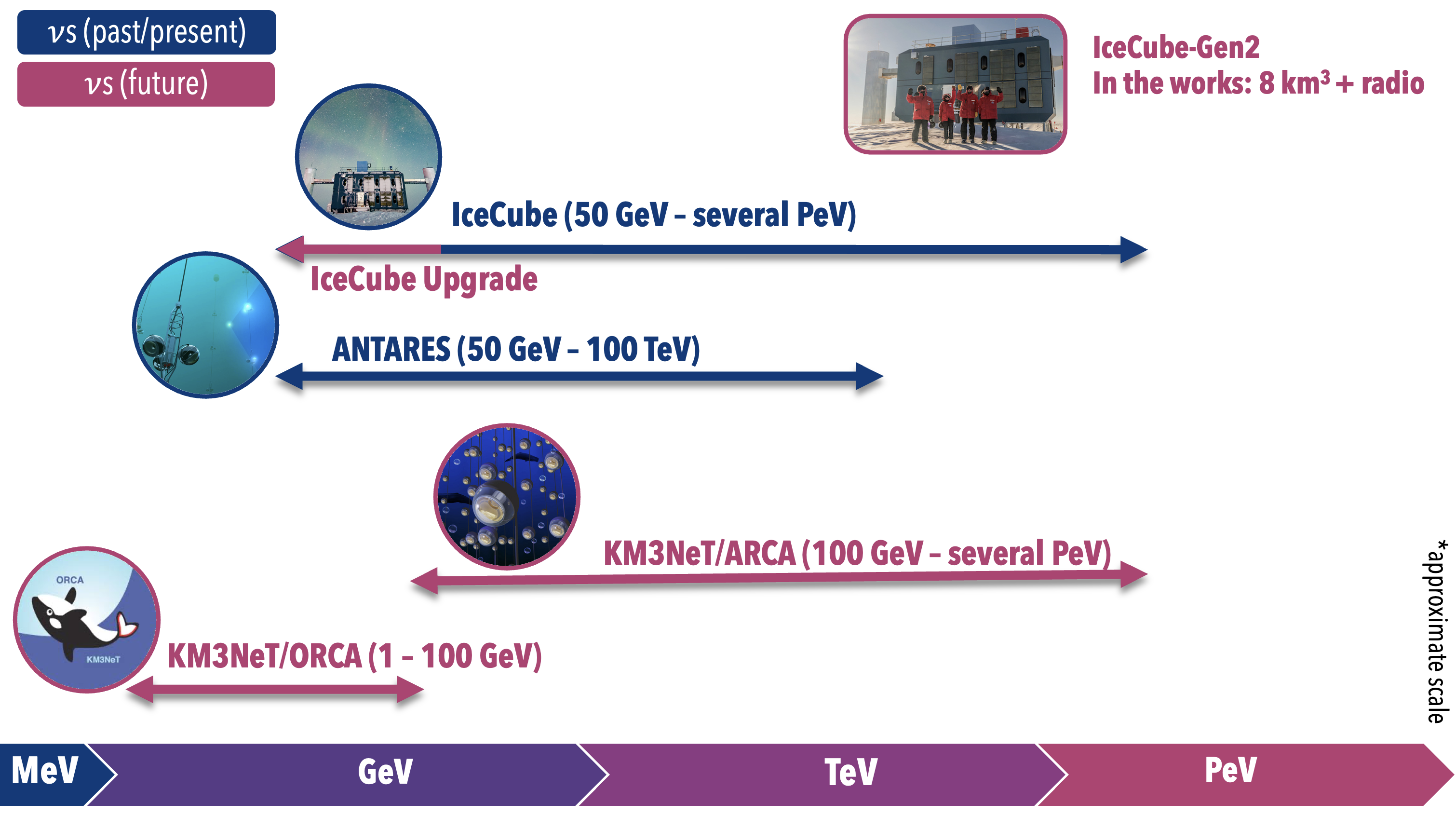}
\caption{\textbf{Neutrino observatories in the GeV–PeV regime.} Current detectors (blue) and future extensions (magenta) provide complementary sky coverage and energy reach.}
\label{fig:neutrinos}
\end{figure}

At ICRC~2025, nearly every astrophysical target class was examined across multiple messengers---$\gamma$-rays, neutrinos, cosmic-ray antimatter, and radio emission---turning what were once single-channel searches into \textbf{genuine multimessenger studies}. Reflecting this shift, I organize this section by \emph{source class} rather than detection technique: dwarf galaxies, the Galactic Center, galaxy clusters, the Sun, and others, each now constrained through complementary observations across multiple messengers. Before turning to specific sources, however, I first highlight the observatories present at ICRC~2025, which illustrate the large part of the current experimental landscape of indirect detection. Figures~\ref{fig:gamma-rays} and~\ref{fig:neutrinos} summarize the $\gamma$-ray and neutrino facilities discussed at the conference, indicating their approximate energy coverage and complementarity.

\subsection{The Sun}

Dark matter gravitationally captured by the Sun through elastic scattering may accumulate and annihilate in its core, producing high-energy neutrinos that escape to Earth. This channel probes the spin-dependent WIMP–proton cross section and complements direct detection for masses above $\sim$10~GeV.

At ICRC~2025, \textbf{IceCube} presented its most sensitive all-flavor Solar dark matter search to date, based on ten years of data and leveraging event selections originally optimized using convolutional neural networks~\cite{IceCube:2025solar}. The analysis covers WIMP masses from 20~GeV to 10~TeV across multiple annihilation channels, achieving world-leading sensitivity over this entire mass range for most channels considered. The limits reach approximately $10^{-41}$ to $10^{-42}$~cm$^2$ for spin-dependent cross sections at 100~GeV---roughly 40 times more stringent than previous IceCube results, particularly benefiting from full treatment of electroweak corrections at high masses. The forthcoming IceCube Upgrade aims to extend this reach to lower masses through improved angular resolution and energy thresholds down to 1--500~GeV.

\textbf{ANTARES} reported final limits using its complete 2007--2022 dataset, concluding more than a decade of operation as the first underwater neutrino telescope~\cite{ANTARES:2025solar}. The analysis uses three independent reconstruction techniques---among them a newly developed machine-learning algorithm optimized for low-energy events---and included all neutrino flavors. The results are consistent with a null signal and achieve competitive sensitivity above 100~GeV, surpassing direct detection constraints in the spin-dependent regime over this mass range. Its successor, the \textbf{KM3NeT/ORCA} array, now under deployment, is expected to continue this program with improved low-energy sensitivity and full flavor reconstruction, directly expanding upon ANTARES's legacy.

A persistent limitation in these searches arises from \emph{solar atmospheric neutrinos}---a flux produced by cosmic-ray interactions in the upper solar atmosphere---which sets a natural sensitivity floor. IceCube's TeV-scale sensitivity is already \textbf{approaching this boundary}, implying that next-generation detectors such as IceCube-Gen2 and KM3NeT will soon probe this irreducible background directly.

At $\gamma$-ray energies, complementary searches target \emph{secluded dark matter} scenarios, where annihilation produces long-lived mediators that decay outside the Solar surface. In these models, the same process generating neutrinos within the Sun yields $\gamma$-rays just beyond it---linking the two messengers as probes of a common underlying physical mechanism. \textbf{The Southern Wide-field Gamma-ray Observatory
(SWGO)} presented updated projections using its latest instrument response functions~\cite{SWGO:2025sun}, indicating sensitivity to spin-dependent cross sections down to approximately $10^{-46}$~cm$^2$ for masses below 5~TeV---an order of magnitude improvement over current indirect limits.


\subsection{The Galactic Center}

The Galactic Center (GC) remains one of the most promising yet challenging regions for indirect dark matter searches. Its high dark matter density enhances the expected annihilation signal, but the region is also dominated by complex astrophysical foregrounds, including the diffuse emission from cosmic-ray interactions, dense molecular gas, and unresolved sources. The long-debated GeV Galactic Center Excess (GCE) observed by \textit{Fermi}-LAT has been largely reinterpreted in terms of millisecond pulsars, yet the persistence of model-dependent residuals continues to motivate a broad experimental effort across messengers and energy ranges.

\begin{figure}[t!]
    \centering
    \includegraphics[width=0.95\textwidth]{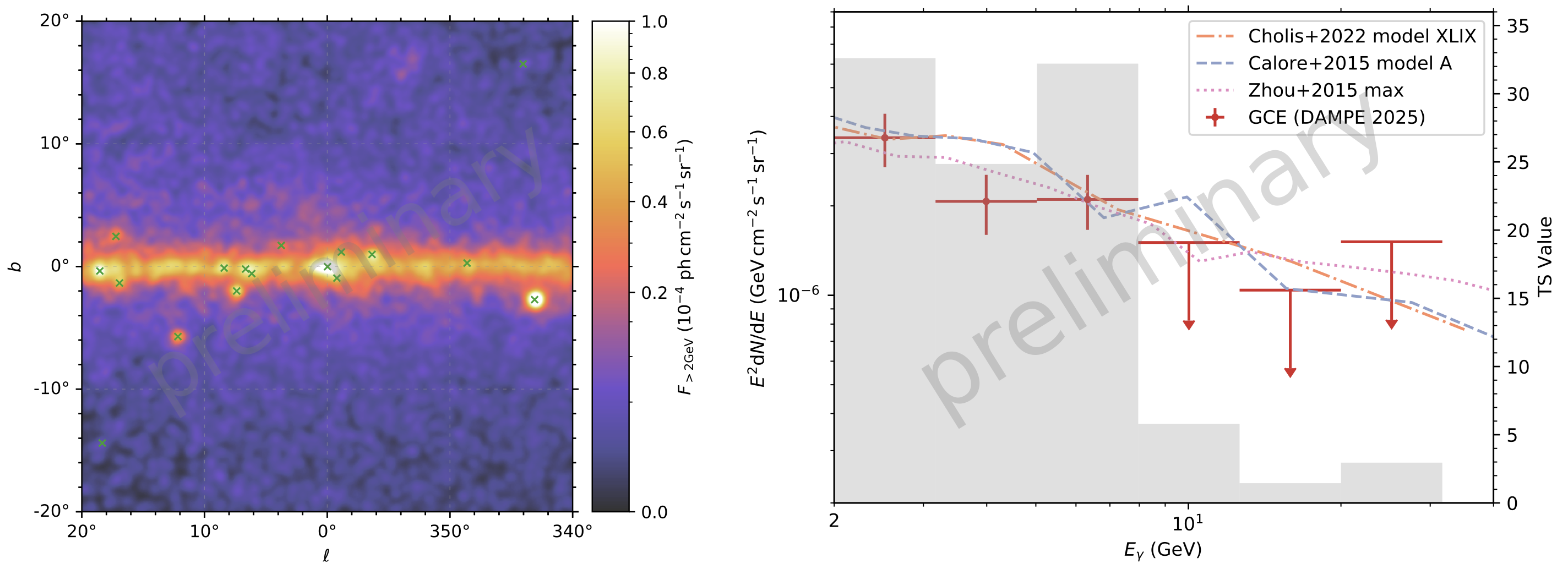}
    \vspace{0.7em}
    \includegraphics[width=0.95\textwidth]{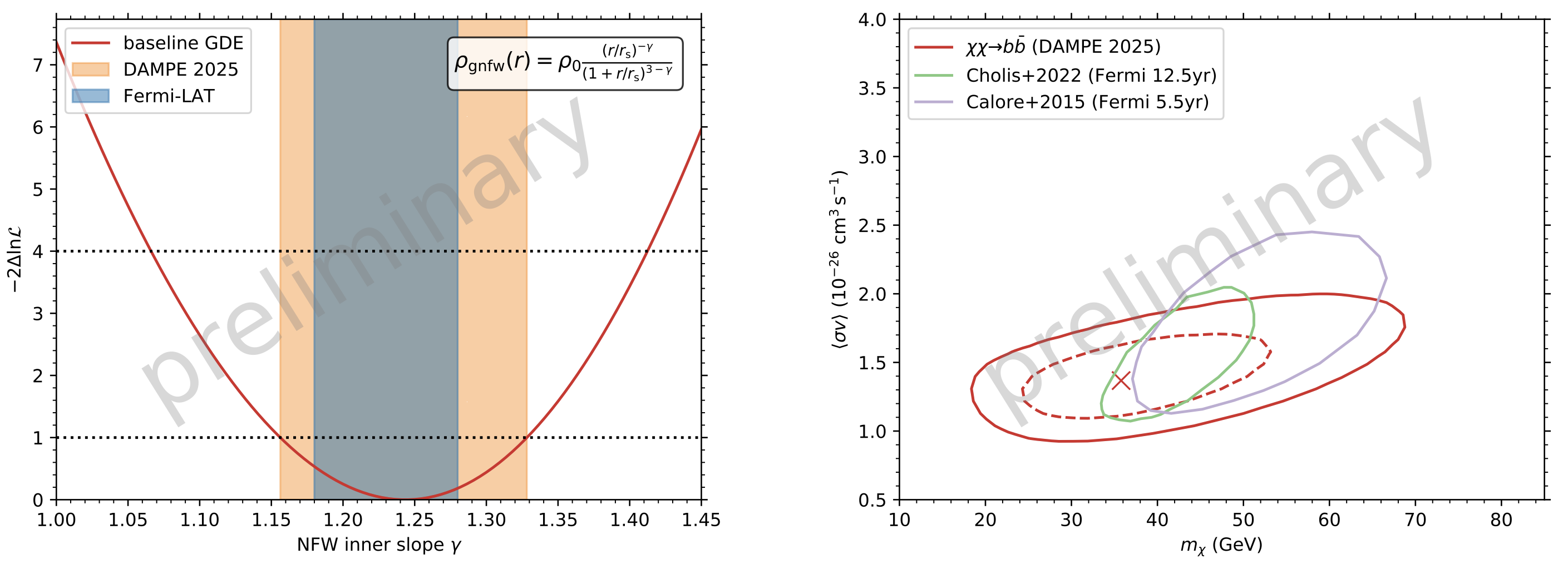}
    \caption{
    \textbf{DAMPE analysis of the GCE.} 
    \textit{Top:} Integrated flux map of the GC region above 2~GeV (left) and the corresponding spectral energy distribution (right). DAMPE reproduces the spectral shape and normalization previously observed by \textit{Fermi}-LAT, providing the first independent confirmation of the excess in $\gamma$-rays. \textit{Bottom:} Likelihood profile for the generalized NFW (gNFW) inner slope $\gamma$ (left) and preferred dark matter parameters for the $\chi\chi\rightarrow b\bar{b}$ channel (right), derived from 8.5 years of DAMPE data. DAMPE countours agree with previous \textit{Fermi}-LAT results (Cholis \textit{et al.}~2022, Calore \textit{et al.}~2015), yielding consistent slopes ($\gamma\sim1.2$) and cross sections ($\langle\sigma v\rangle \sim 1.4\times10^{-26}$ cm$^3$ s$^{-1}$). The slightly lower preferred $m_\chi\sim35$~GeV compared to \textit{Fermi}-LAT likely reflects DAMPE's limited sensitivity at higher energies, which can make it less constraining toward lower-mass dark matter models. Adapted from \cite{Shen:2025}.}
    \label{fig:DAMPE_GC}
\end{figure}

At ICRC~2025, the \textbf{first Large-Sized Telescope (LST-1) of the Cherenkov Telescope Array Observatory (CTAO)} presented updated sensitivity studies for $\gamma$-ray line searches toward the GC \cite{Abhishek:2025}. The analysis targets annihilation channels producing monochromatic photons ($\chi\chi\rightarrow\gamma\gamma$ and $\chi\chi\rightarrow\gamma Z$) in the 1--100~TeV range. Large zenith angle observations enhance the effective area while preserving low thresholds, enabling sensitivity improvements over previous MAGIC and H.E.S.S. results in the multi-TeV regime. The LST-1 also emphasized a detailed treatment of systematics in background modeling---an important step toward realizing the full CTAO potential for spectral line searches once the full array is operational.

Complementary results now come from neutrino observatories probing the same region. \textbf{IceCube}, using DeepCore and forthcoming Upgrade extensions, presented a dedicated GC search across the GeV--TeV range \cite{Chau:2025}. Although neutrino channels remain less sensitive than $\gamma$-rays for leptonic annihilation modes, they offer a clean probe of hadronic final states and nonstandard topologies. The analysis of 9.3 years of DeepCore data set world-leading limits on neutrino-line signals and provided the first projections for the IceCube Upgrade, which is expected to improve sensitivity below 20~GeV by an order of magnitude.

The \textbf{KM3NeT} collaboration expanded this program to the opposite hemisphere, combining ORCA (optimized for 1--100~GeV) and ARCA (extending to 10~PeV) \cite{BariegoQuintana:2025}. Their joint likelihood framework---tested on configurations from 6 to 230 detector lines---delivers nearly continuous sky coverage of the inner Galaxy. Even with partial configurations, KM3NeT reaches competitive limits above 10~TeV and demonstrates that multi-detector synergy is rapidly becoming standard practice in neutrino-based DM searches. 

Looking ahead, the \textbf{ Hyper-Kamiokande} collaboration, presenting in the neutrino track, outlined prospects for detecting the astrophysical neutrino flux and for dark matter annihilation searches toward the GC, showing potential sensitivity to the $\chi\chi \to \nu\bar{\nu}$ channel in the 10--$10^4$~GeV range and complementarity with IceCube and KM3NeT \cite{RamosCason:2025}.

Finally, a distinctive and creative approach linked cosmic-ray ionization in the Central Molecular Zone (CMZ) to potential MeV-scale dark matter interactions \cite{DeLaTorreLuque:2025}. This analysis connected an otherwise unexplained astrophysical anomaly---the elevated ionization rate of the CMZ---with low-mass dark matter capable of producing secondary electrons and positrons. Though highly model-dependent, it exemplified the field's growing willingness to explore noncanonical mass regimes and multi-wavelength observables.

Outside the official dark matter track at ICRC~2025, yet nonetheless one of my favorite moments of the conference came from \textbf{DAMPE}. Using 8.5~years of photon data, the collaboration presented \textbf{the first independent confirmation of the well-established Galactic Center Excess} (GCE) in GeV $\gamma$-rays (7.4$\sigma$ significance)---sixteen years after it was first observed with \textit{Fermi}-LAT \cite{Shen:2025}. While there is no doubt that the excess is genuine (\textit{Fermi}’s result has been extensively scrutinized)---it is deeply gratifying to see it re-emerge in data from an entirely independent instrument, with distinct detection techniques and systematic uncertainties. \textbf{The GCE remains well and alive, still awaiting resolution.}

\subsection{The Galactic Halo}

The Milky Way halo remains the most direct yet diffuse target for dark matter searches. While its total mass distribution sets the scale for all Galactic analyses, its substructure, density profile, and annihilation morphology shape both theoretical predictions and instrumental strategies.
At ICRC~2025, contributions spanned from high-resolution simulations of halo substructure to deep observational searches for spectral features and antimatter signatures, underscoring the halo's role as a testable component of the Galactic dark matter distribution.

High resolution \textbf{$N$-body simulations} resolved subhalo tidal tracks with particle counts exceeding $10^7$, providing a detailed picture of how gravitational stripping shapes the surviving substructure population \cite{AguirreSantaella:2025}. Complementary magneto-hydrodynamic simulations that include gas and magnetic feedback showed that baryonic processes can significantly suppress low-mass subhalos, leading to more accurate predictions for the number and brightness of dark satellites potentially observable with \textit{Fermi}-LAT \cite{Porras:2025a}.

At smaller scales, late-time annihilations in oscillating asymmetric dark matter (OADM) models were explored as a possible dynamical solution to the core–cusp problem \cite{Mamprim:2025}. In this scenario, a small Majorana mass term reactivates particle-antiparticle annihilations during galaxy evolution, gradually transforming cuspy NFW profiles into constant-density cores.
Fits to dwarf-galaxy rotation curves indicate that this mechanism can reproduce observed density cores for sub-GeV particle masses and cross sections of order $\sigma/m_\chi \sim 0.06$~cm$^2$,g$^{-1}$, providing a particle physics alternative to purely baryonic feedback. From an observational standpoint, spectral searches for monochromatic and box-like $\gamma$-ray features found no evidence of dark matter annihilation. \textbf{DAMPE}'s nine-year dataset and the updated 15.5-year \textbf{\textit{Fermi}-LAT} analysis both reported null results, with the latter improving previous limits by up to two orders of magnitude through refined background modeling below 10~GeV \cite{Frieden:2025, Giliberti:2025}.

\begin{figure}[t]
\centering
\begin{minipage}[t]{0.38\textwidth}
    \centering
    \includegraphics[width=\textwidth]{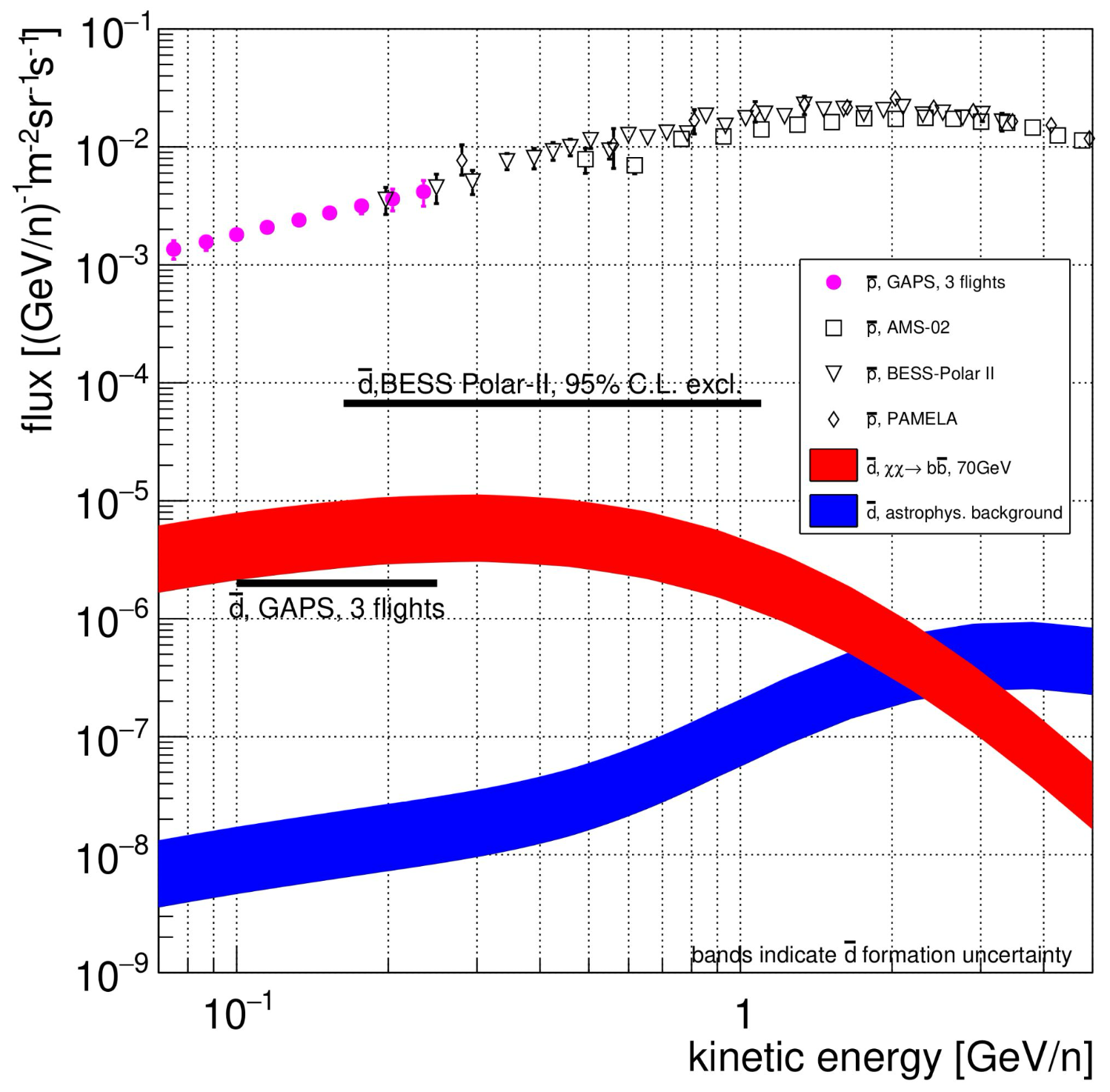}
    \vspace{-0.4em}
\end{minipage}
\hfill
\begin{minipage}[t]{0.6\textwidth}
    \centering
    \includegraphics[width=\textwidth]{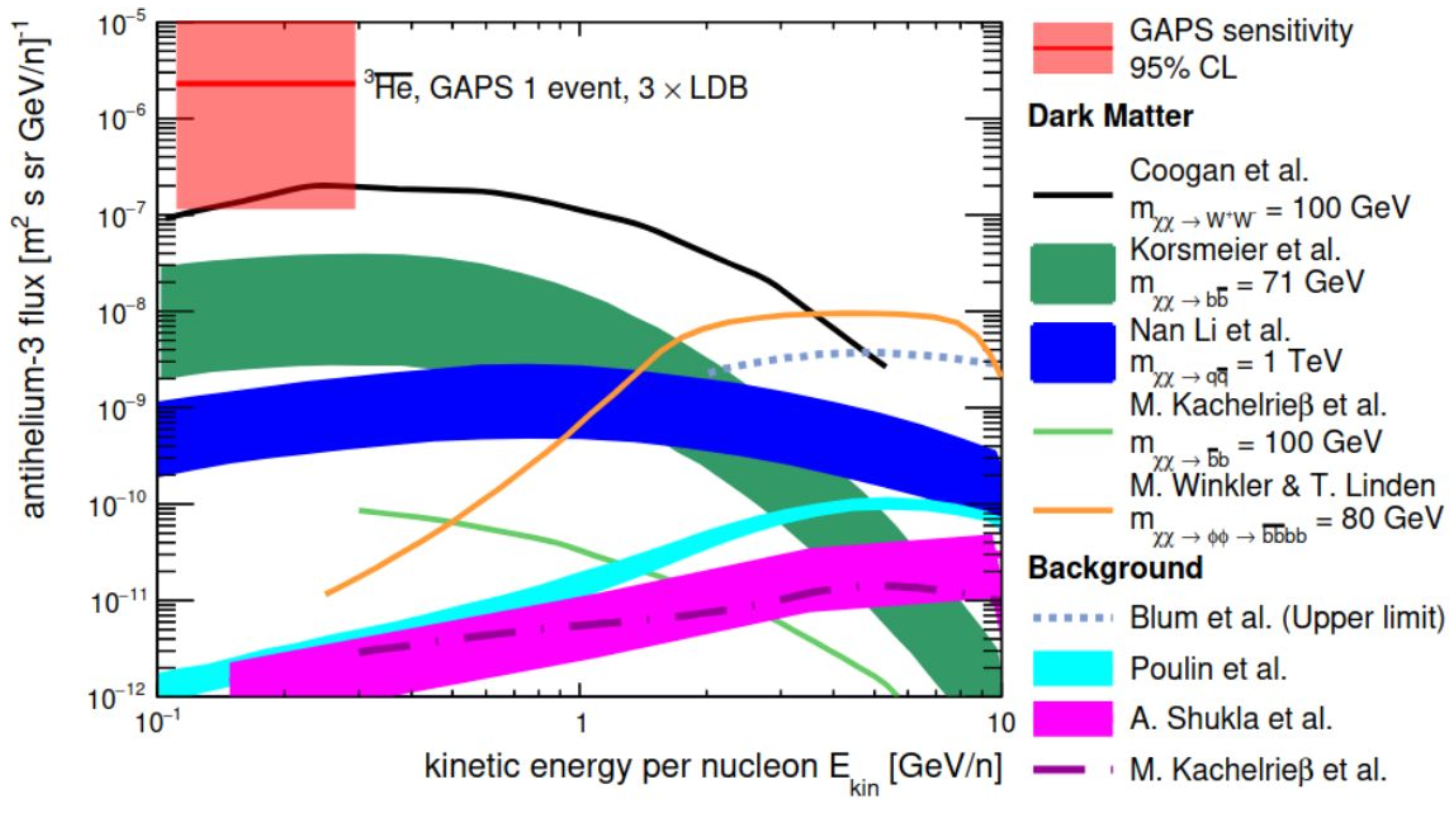}
    \vspace{-0.4em}
\end{minipage}
\caption{\textbf{Left:} Predicted fluxes of antiprotons and antideuterons from dark matter annihilation compared with astrophysical backgrounds and existing measurements. The GAPS long-duration balloon flights will probe the low-energy regime ($E_{\mathrm{kin}} \lesssim 0.25$~GeV/n), where astrophysical backgrounds sharply decline, enabling nearly background-free searches for antideuterons. 
\textbf{Right:} Projected GAPS sensitivity to antihelium-3 after three long-duration flights (red band, 95\%~CL), compared with expected fluxes from representative dark matter models and astrophysical backgrounds. Adapted from \cite{Stoessl:2025}.}
\label{fig:gaps_antimatter}
\end{figure}

The search for \textbf{cosmic antimatter} stood out at ICRC~2025 as the part of indirect detection most clearly oriented towards \textbf{future experiments}.
These searches target low-energy cosmic antinuclei---antiprotons, antideuterons, and antihelium---where, especially for the latter two, astrophysical backgrounds drop sharply below a few hundred~MeV per nucleon, making even a handful of detected events potentially decisive.
Three complementary efforts defined the state of the field: \textbf{GAPS}, \textbf{GRAMS}, and \textbf{PLASTICAMI}.

The \textbf{General Antiparticle Spectrometer (GAPS)} collaboration presented results from its full-system ground campaign in Antarctica~\cite{Aoyama:2025, Stoessl:2025}.
GAPS employs a novel \emph{exotic-atom technique}, in which low-energy antinuclei are captured by lithium-drifted silicon detectors, de-exciting through x-ray emission before annihilating into pions. This dual signature provides excellent particle identification and background suppression, enabling sensitivity to antideuterons and antihelium at fluxes several orders of magnitude below those accessible to magnetic spectrometers. Although \textit{seven consecutive} launch attempts were \textit{weather-prevented}, the payload remains integrated and flight-ready at McMurdo Station, with its first long-duration Antarctic flight now expected in the 2025/26 season. Once launched, GAPS will extend current limits by one to two orders of magnitude, probing fluxes predicted for $\sim$70~GeV WIMP annihilation and measuring the antiproton spectrum at kinetic energies below 0.25~GeV/nucleon.

The \textbf{GRAMS (Gamma-Ray and AntiMatter Survey)} project advanced a complementary concept based on a dual-purpose liquid-argon TPC~\cite{Aramaki:2025}.
GRAMS aims to bridge the long-standing ''MeV gap`` in $\gamma$-ray astronomy while simultaneously searching for antideuterons and antihelium. A successful engineering flight and J-PARC beam test demonstrated its detector performance, and a NASA-funded prototype mission (\textit{pGRAMS}) is now scheduled for early~2026.
With a modular design scalable to space deployment, GRAMS could ultimately connect MeV $\gamma$-ray observations and low-energy cosmic-ray antimatter measurements within a single instrument class.

Finally, \textbf{PLASTICAMI} proposed a segmented plastic-scintillator tracker optimized for identifying the unique ''two-step`` annihilation topology of stopped antideuterons in hydrogen-rich media~\cite{Ghezzer:2025}.
By timing the delayed annihilation of the surviving antinucleon following the first annihilation, PLASTICAMI can isolate the antideuteron signature against background.
Simulations show that two long-duration balloon flights could reach flux sensitivities of $\Phi_{\bar{d}}\sim2\times10^{-6}$~(m$^2$\,sr\,s\,GeV/n)$^{-1}$ between 100–600~MeV/n, approaching the discovery region for several dark matter annihilation models.

\subsection{Dwarf Spheroidal Galaxies}

Dwarf spheroidal galaxies (dSphs) remain benchmark targets for indirect dark matter searches. Their extreme mass-to-light ratios, combined with minimal astrophysical backgrounds, and kinematically well-constrained dark matter distributions, make them exceptionally well-suited for testing different dark matter models. Unlike the GC, where baryonic emission complicates interpretation, dSphs offer a more controlled environment where systematic uncertainties are dominated by the astrophysical $J$-factor rather than foreground contamination or source confusion. 

A major milestone in the field was the completion of the \emph{first multi-instrument joint-likelihood analysis}, combining data from \textbf{\textit{Fermi}-LAT, HAWC, H.E.S.S., MAGIC, and VERITAS} across twenty dwarf galaxies~\cite{Rico:2025}. First presented at ICRC~2019 and finalized this year, this effort joined five independent likelihood datasets---spanning from 100~MeV to beyond 10~TeV---under consistent statistical and astrophysical assumptions. The analysis explicitly combined systematics and instrument response functions, resulting in continuous coverage across eight decades in energy. Sensitivity is dominated by \textit{Fermi}-LAT at low masses and by the Cherenkov telescopes at higher energies, where the joint framework improves limits by factors of two to three relative to any single instrument. For a 2~TeV WIMP annihilating into $\tau^+\tau^-$, the combined limit reaches $\langle\sigma v\rangle\lesssim3\times10^{-25}$~cm$^3$~s$^{-1}$. More broadly, however, this result \textbf{sets a precedent for cross-facility analyses}, providing a methodological foundation for future joint searches that may also include neutrino observatories.
\begin{figure}[t]
\centering
\includegraphics[width=0.48\textwidth]{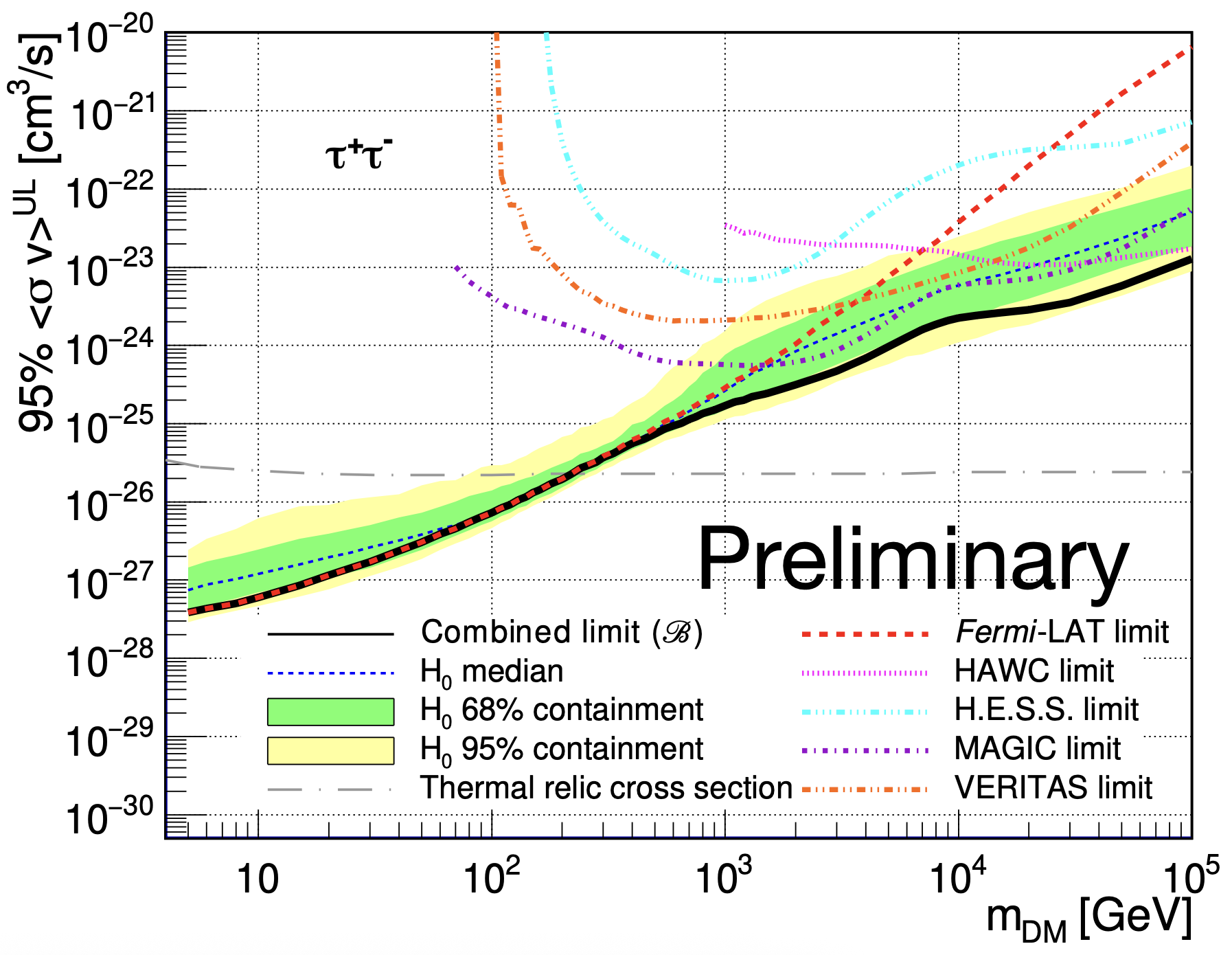}
\includegraphics[width=0.48\textwidth]{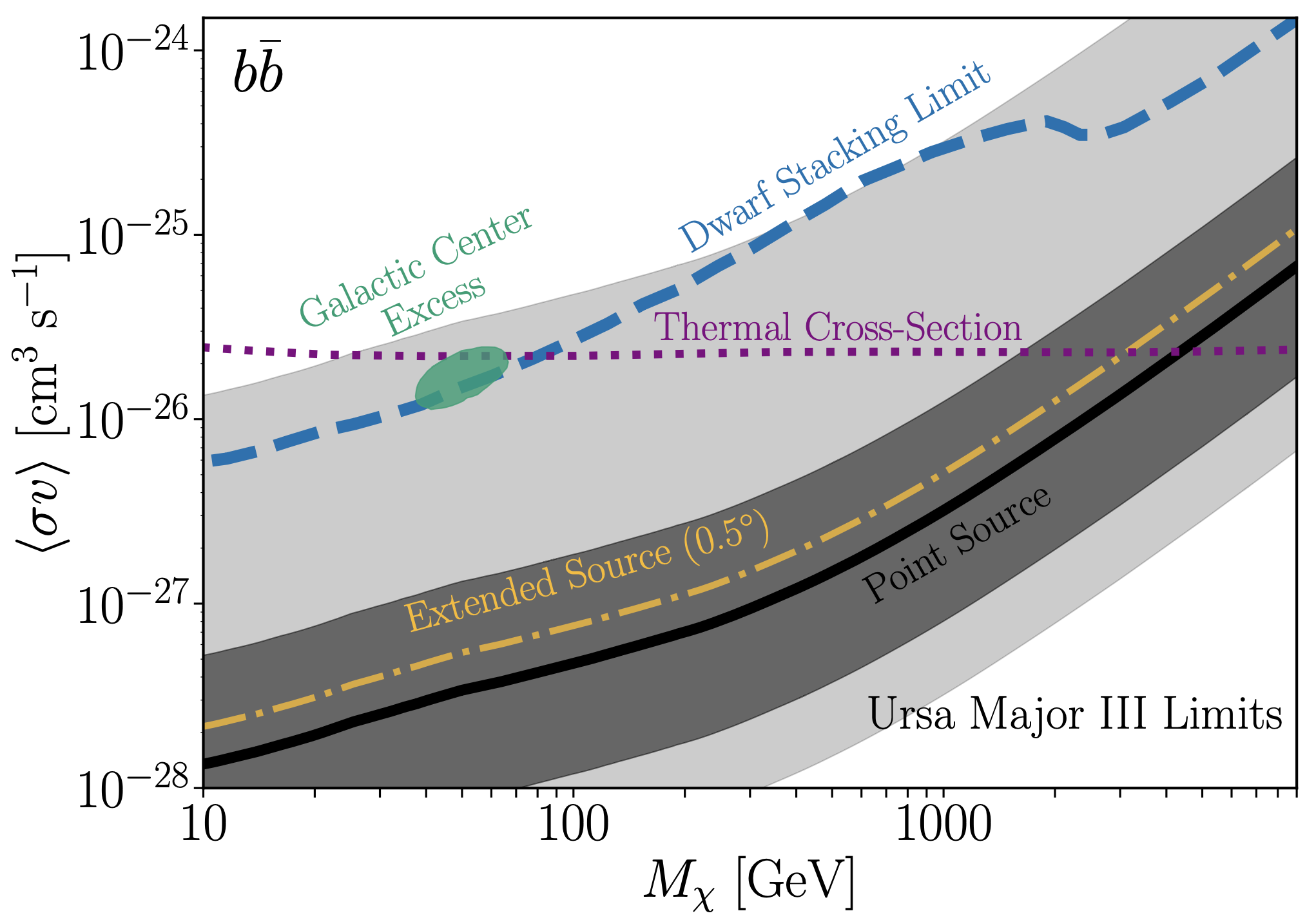}
\caption{\textbf{Two complementary paths in dwarf-galaxy dark matter searches.}
Left: The joint \textit{Fermi}-LAT, HAWC, H.E.S.S., MAGIC, and VERITAS analysis~\cite{Rico:2025} 
combines five likelihoods across eight decades in energy, exemplifying the integration of instruments and systematics into a unified statistical framework.
Right: \textit{Fermi}-LAT limits for the ultra-faint galaxy \textbf{Ursa Major~III}~\cite{McGrath:2025,Crnogorcevic:2025}, 
illustrating the complementary precision approach that targets individual, extreme systems with high inferred dark matter densities.}
\label{fig:dSph_combined}
\end{figure}

Complementing this, \textbf{VERITAS} presented their results for the ultra-faint galaxy \textbf{Ursa Major~III} in combination with \textit{Fermi}-LAT data~\cite{McGrath:2025}. Rather than extending exposure or joining likelihoods, the analysis focused on a single, extreme dark matter system with a high $J$-factor, exploring both canonical WIMP channels and specific electroweak scenarios such as wino and quintuplet dark matter. Ultra-faint dwarfs like Ursa Major~III are becoming increasingly common and better characterized through facilities such as Rubin~LSST, enabling targeted follow-up of individual systems with high inferred dark matter densities. Although the confirmation and dynamical characterization of such faint systems remain challenging---and some may ultimately prove less dark matter dominated than currently assumed---their discovery continues to expand the range of viable targets for indirect searches, allowing us to turn towards precision studies of the faintest and most dark matter dominated systems. 

In the radio domain, \textbf{MeerKAT} presented the deepest interferometric dark matter search in a dwarf galaxy to date~\cite{Semane:2025}, targeting Reticulum~II in the UHF band. The study detected no significant emission but achieved sensitivities of 9--14~$\mu$Jy/beam, improving on previous ATCA constraints by up to an order of magnitude. For WIMP masses between 20--100~GeV annihilating into $b\bar{b}$, the analysis excluded cross-sections above $\sim$10$^{-26}$~cm$^3$s$^{-1}$. These results represent the most stringent radio constraints obtained for a dwarf galaxy, demonstrating MeerKAT's capability as a precision instrument for indirect dark matter searches. The forthcoming Square Kilometre Array (SKA) is expected to extend this reach by one to two orders of magnitude, potentially probing the thermal relic scale in the radio band.

At the highest energies, \textbf{IceCube} presented its first complete dSph analysis using the muon-track channel~\cite{IceCube:2025dSph}. Based on 10.4~years of data toward fifteen northern-sky dwarfs, the study distinguishes neutrino-flavor signatures from dark-matter-induced fluxes and achieves competitive sensitivity above 100~TeV. The stacking approach and flavor-resolved modeling make this the first neutrino analysis of its kind, demonstrating clear complementarity with $\gamma$-ray observatories in probing heavy dark matter scenarios.

\textbf{Theoretical modeling efforts} presented at ICRC~2025 explored prompt and secondary emission in transition dwarf galaxies---systems that have recently ceased star formation---to distinguish dark-matter–induced signals from astrophysical backgrounds~\cite{Aravinthan:2025}. Using \texttt{CRPropa}-based simulations, these studies showed that synchrotron and inverse-Compton components imprint distinct spatial and spectral morphologies, underscoring the diagnostic power of multi-wavelength structure in future searches.

Taken together, these results establish dSphs as \textbf{genuinely \emph{multimessenger} dark matter targets}. The joint likelihood $\gamma$-ray framework now combines space- and ground-based observations across eight decades in energy, while radio and neutrino analyses extend sensitivity to complementary regimes, probing secondary emission and heavy-mass scenarios inaccessible to photons alone. At the same time, the field is branching along \textbf{two complementary paths}: one toward integration, where coordinated multi-instrument analyses redefine the statistical frontier; and one toward precision, where ultra-faint dwarfs such as Ursa Major~III invite deep, system-specific consideration. As new systems are discovered and their dark matter content better constrained through deep spectroscopic and photometric surveys, the population of dwarf galaxies continues to expand both in number and in diversity, emerging as one of the most promising targets for testing dark matter interactions in the coming decade.

\subsection{Extragalactic Sources}
Beyond the Milky Way, galaxy clusters and active galactic nuclei (AGN) provide complementary laboratories for dark matter searches. Their large gravitational potentials, extended dark matter halos, and (often) well-characterized intracluster media allow indirect searches to test both annihilation and decay scenarios over large spatial scales. At ICRC~2025, extragalactic searches reflected\textbf{ two trends}: (i) refined analyses of nearby clusters as extended sources of potential annihilation signatures, and (ii) a broad diversification of probes beyond the canonical WIMP, connecting $\gamma$-ray, radio, and multimessenger signatures across different mass regimes.

\noindent\textbf{Galaxy clusters.} 
At radio wavelengths, \textbf{MeerKAT} has established a new benchmark for indirect dark matter searches through the \textit{Galaxy Cluster Legacy Survey}~\cite{ICRC2025_MeerKAT}. The program analyzed 115 clusters using an image-domain pipeline capable of isolating diffuse synchrotron emission from compact-source contamination. The resulting upper limits on leptonic annihilation channels are amongst the most stringent to date, showing the power of radio interferometry as a complementary probe to $\gamma$-ray observations. Furthermore, this work introduced a computationally efficient alternative to traditional visibility-based analyses through injecting dark matter templates directly into processed images. The method reduces processing time from 82~hours to roughly 3~minutes per field on a single node---orders of magnitude faster---resolving a critical scalability challenge for next-generation surveys such as the SKA. Complementary work in the $\gamma$-ray track included a dedicated \textbf{H.E.S.S.} search for dark matter signatures in the Virgo Cluster, based on 200~hours of observations and modeling both annihilation and decay channels with substructure effects \cite{Bradascio:2025}.

A separate \textbf{\textit{Fermi}--LAT} analysis reported a tentative 43~GeV $\gamma$-ray line feature in a stacked study of nearby clusters, including Virgo, Fornax, and Ophiuchus~\cite{ICRC2025_LAT_43GeV}. The signal reaches a post-trial significance of 3.7$\sigma$ for a 13-cluster sample, rising to 4.3$\sigma$ when restricted to the three highest-$J$ systems. Multiple consistency checks---testing the line’s monochromatic width, spatial extent within the virial radius, and Earth-limb control regions were performed, but instrumental origins remain under investigation. This feature appears absent in the inner Galaxy, creating tension with standard annihilation interpretations and suggesting either more complex dark sector dynamics or unaccounted-for systematics. Further instrumentation effect analysis and/or verification by CTAO or HAWC will be essential before any dark matter interpretation can be sustained. 

\noindent\textbf{Jetted Sources and ALP Searches.}
ICRC~2025 featured a \textbf{strong focus on non-WIMP scenarios}, particularly ALPs and their imprints in extragalactic environments. \textbf{HAWC} presented two complementary studies: one analyzing TeV blazars (VER~J0521+211, 1ES~0229+200, and PG~1553+113) with updated data, extending existing constraints in the neV-scale ALP parameter space~\cite{ICRC2025_HAWC_Blazars}; and another based on seven years of observations of M87, employing a Monte Carlo likelihood framework that explicitly circumvents Wilks’ theorem assumptions~\cite{ICRC2025_HAWC_M87}. Both analyses yield new limits on the photon-ALP coupling strength and illustrate the growing methodological maturity of astrophysical ALP searches.
Complementary simulations for the upcoming \textbf{SWGO} indicate that five years of observations of Centaurus~A could further refine ALP constraints by detecting—or excluding—spectral distortions induced by photon–ALP mixing~\cite{ICRC2025_SWGO_ALPs}.

A separate effort modeled the interplay between dark matter and cosmic rays in relativistic jets, introducing a multi-zone framework that accounts for jet geometry and kinematics when constraining cosmic-ray–dark-matter (CR–DM) elastic and inelastic interactions~\cite{ICRC2025_Jets_DM}. Using Markarian~421 as a representative case, the study found that jet dynamics significantly affect expected signatures, underscoring the need for physically consistent modeling of such environments.
At higher energies, theoretical work examined the impact of photophilic ALPs in $\gamma$-ray bursts (GRBs), showing that MeV--GeV-mass ALPs could suppress GRB emission---an effect we do not observe---thereby imposing strong limits on heavy ALPs coupled to photons~\cite{ICRC2025_GRB_ALPs}.
Finally, models connecting neutrino production to dark matter interactions in the dense spikes surrounding supermassive black holes explored how such regions could contribute to the observed neutrino fluxes from active galaxies like NGC~1068, bridging multimessenger observations across neutrinos and $\gamma$~rays~\cite{ICRC2025_DM_Spikes}.

Rather than adding more upper limits, the extragalactic program at ICRC~2025 broadened what a dark matter search can be. Galaxy clusters showcased advances in computational pipelines and synchrotron modeling, blazars and radio galaxies are studied as sites for ALP–photon mixing, and jets and GRB fireballs emerged as probes of dark sector interactions across the lightest and heaviest mass scales. In preparation for upcoming facilities such as \textbf{CTAO, SWGO, and SKA,} this integration of astrophysical data handling and dark matter phenomenology promises a more direct translation of observational signatures into particle physics insights in the coming decade.

\subsection{Large Scale Structure}

Large Scale Structure (LSS) entered the dark matter program at ICRC~2025 less as a background assumption and more as a source of testable constraints. Across several contributions, cosmological modeling and early-Universe dynamics were directly linked to the interpretation of astrophysical searches.

\noindent\textbf{Primordial black holes (PBHs)} remain the most direct bridge between cosmology and astroparticle physics. The \textbf{LHAASO} collaboration conducted an all-sky search for individual PBH burst events using 3.4~years of WCDA data, finding no significant excess and setting a new upper limit on the local burst rate density of $\rho_{\mathrm{PBH}}<181~\mathrm{pc}^{-3},\mathrm{yr}^{-1}$ at 99\%~C.L., the most stringent bound to date \cite{Jiang:2025}. While still consistent with purely cosmological populations, these limits begin to challenge models in which PBHs constitute a significant fraction of the dark matter density. Complementary constraints came from charged-particle channels. Analyses using updated \textbf{AMS-02} antiproton data and refined Galactic propagation models revisited antinuclei production from PBH evaporation \cite{Stefanuto:2025}. Incorporating lognormal PBH mass distributions and improved coalescence modeling for antideuteron formation, these studies showed that cosmic-ray data already exclude large portions of PBH parameter space, comparable to current $\gamma$-ray limits. In effect, the cosmic-ray background itself now functions as a probe of early-Universe structure formation.

Another study modeled the cosmic background from optical to x-ray energies, incorporating ALP decays on LSS–averaged extragalactic background light templates~\cite{Porras:2025b}. By comparing with recent cosmic-background measurements---including the \textbf{New Horizons} sky-brightness excess---the study derives improved limits on the ALP–photon coupling, reaching the QCD axion band.

At smaller scales, hydrodynamical simulations such as \textbf{EAGLE} are directly informing indirect searches. Analyses combining LSS-based halo realizations with dark matter spike models around intermediate mass black holes showed how realistic substructure modifies the expected $\gamma$-ray flux in \textbf{H.E.S.S.}\ data \cite{Vecchi:2025a, Vecchi:2025b}. No excess consistent with annihilation was found, but the resulting limits are amongst the strongest obtained for intermediate-mass systems and provide an astrophysically grounded bridge between cosmological simulations and indirect $\gamma$-ray searches.

Finally, \textbf{IceCube} extended these cosmological connections to relic particle searches, presenting a twelve-year analysis of sub-relativistic magnetic monopoles that improves previous flux limits by up to three orders of magnitude \cite{Haussler:2025}.
\subsection{Macroscopic Dark Matter and Nuclearites}
In this subsection, I depart slightly from the source-based organization used elsewhere and turn to dark matter candidates in the macroscopic regime. Dark matter may also exist in composite rather than elementary form, appearing as dense, massive objects that interact primarily through geometry and gravity. Such candidates---including strange-quark matter nuggets, nuclearites, and other compact macroscopic dark objects---span mass range from $10^{18}$ to $10^{32}$~GeV and could produce meteor-like luminous tracks or measurable acoustic signatures as they traverse the atmosphere or solid media.

The \textbf{DIMS (Dark Matter and Interstellar Meteoroid Study)} experiment employs a network of high-sensitivity CMOS cameras at the Telescope Array site in Utah to identify slow, faint luminous events consistent with nuclearite passage~\cite{Kajino:2023}. Over 450~hours of moonless observations have been collected, yielding tens of thousands of meteor detections and establishing a framework for distinguishing macroscopic dark matter candidates from conventional meteoroids through their higher velocities, lower altitudes, and potential upward trajectories.

In space, the \textbf{SQM-ISS} detector~\cite{Plebaniak:2025}---approved for installation on the International Space Station---combines scintillator and piezoelectric layers to detect both ionization and mechanical energy deposition from slow, massive objects such as strange-quark matter, Q-balls, or PBHs. Operating in a galactic-velocity regime ($v\sim220$~km s$^{-1}$), the instrument aims to provide the first orbital constraints on macroscopic compact dark matter, with projected sensitivities several orders of magnitude beyond existing ground-based limits. 

\section{Conclusions: The Dark Matters at ICRC~2025 (or, my 2\textcent)}
\label{sec:concl}

We have reached a mature phase in dark matter searches\footnote{A quiet tap on the shoulder reminding us how far the field has come.}. The field has moved beyond isolated anomalies towards disciplined, reproducible, and statistically rigorous analyses. What began as the WIMP paradigm has evolved into a broader framework encompassing axions, ultralight fields, sub-GeV candidates, and macroscopic relics. These efforts are no longer parallel pursuits but parts of a \textbf{coordinated multimessenger program} built on common analysis standards and shared data infrastructure.

\subsection*{Key Developments}
\noindent \textbf{\underline{The neutrino floor is here.}} Direct detection experiments have reached the background set by Solar and atmospheric neutrinos, indicated by the CE$\nu$NS measurements in XENONnT and PandaX-4T. Future gains will require both larger exposures and a deeper control of irreducible backgrounds---understanding of backgrounds, not scale alone, will determine the next big result.
\smallskip
\\
\noindent\textbf{\underline{Joint analysis frameworks define the state of the art.}} Cross-instrument likelihoods have replaced single-experiment limits as the measure of progress. Coordinated statistical frameworks---shared priors, consistent background models, and open likelihoods---now drive sensitivity gains as much as raw exposure. A host of upgraded and forthcoming facilities---DarkSide-20k, SuperCDMS, DarkSPHERE, SABRE South, IceCube Upgrade and Gen-2, KM3NeT, SWGO, CTAO, GAPS, GRAMS, and the SKA---will soon expand our observational reach, providing us with an unprecedented volume and diversity of data to be analyzed within cross-instrument frameworks that now underpin the field.
\smallskip
\\
\noindent\textbf{\underline{The candidate space has broadened and matured.}}
Axions, ultralight fields, sub-GeV states, and macroscopic relics are now pursued through coordinated laboratory, astrophysical, and cosmological efforts. What began as separate niche searches has become a coherent strategy spanning more than eighty orders of magnitude in mass.

\subsubsection*{Challenges Ahead}
Still, realism tempers my optimism. Many of these advances depend on sustained support for mid-scale experiments, theory-instrument integration, and continuity in space-based $\gamma$-ray coverage---\textbf{none} of which are guaranteed. Without a successor to \textit{Fermi}-LAT, or a coordinated MeV mission, the multimessenger bridge risks becoming \textit{one-sided}. ICRC~2025 showed that the field's limitations are increasingly structural rather than technical. The decisive progress in the coming decade is likely to depend on how effectively the community coordinates analyses across experiments, instruments, and messengers---turning parallel efforts into a coherent framework rather than a collection of incremental sensitivity gains.

Where does that leave us? Four decades after the modern search began, the identity of dark matter remains unknown---but some of its absence is very well-constrained. The field's maturity is in itself an achievement: a demonstration that precision, patience, and shared infrastructure can turn even the null results into scientific progress. Whether we find dark matter in two years, ten, or fifty, I believe the discovery will trace its lineage to \textit{this very moment}---when dark matter physics learned to see more by combining what it already has, and to nurture---as exemplified at ICRC 2025---the community that carries it forward.

\section*{Acknowledgments}

I thank the organizers of ICRC 2025 for the opportunity to serve as Dark Matter rapporteur and the 78 contributors whose presentations and proceedings informed this summary. Special thanks to Tim Linden, Silvia Manconi, Michael Korsmeier, Ludwig Neste, Carlos Blanco, Isabelle John, Pedro De la Torre Luque, Thong Nguyen, and Alexandre Adler for their insightful discussions that shaped this rapportour paper. I additionally acknowledge the use of OpenAI's ChatGPT to assist with
editing and proofreading of the manuscript. My work is supported by the Swedish Research Council under contract 2022-04283 and the Swedish National Space Agency under contract 117/19.

\end{document}